\documentclass[final]{ws-mpla}
\pdfoutput=1
\usepackage[super,comma,sort&compress]{natbib}
\usepackage{graphicx}
\usepackage{ifdraft}
\usepackage{ifpdf}
\ifpdf
	\DeclareGraphicsExtensions{.png, .jpg}
\else
	\DeclareGraphicsExtensions{.eps, .png, .jpg}
\fi	

\ifdraft{\let\trimmarks\cropmarks}{\let\trimmarks}

\begin{document}

\markboth{M. Zemp}{The Structure of Cold Dark Matter Halos}

\title{THE STRUCTURE OF COLD DARK MATTER HALOS: RECENT INSIGHTS FROM HIGH RESOLUTION SIMULATIONS\footnote{This invited review is based on talks given by the author in 2008-2009.}}

\author{\footnotesize MARCEL ZEMP}

\address{University of Michigan, Department of Astronomy\\
500 Church Street, Ann Arbor, MI 48109-1042, USA\\
mzemp@umich.edu}

\maketitle

\pub{Received 2. September 2009}{}

\begin{abstract}
We review results from recent high resolution cosmological structure formation simulations, namely the Via Lactea I \& II and GHALO projects. These simulations study the formation of Milky Way sized objects within a cosmological framework. We discuss the general properties of cold dark matter halos at redshift $z=0$ and focus on new insights into the structure of halos we got due to the unprecedented high resolution in these simulations.
\end{abstract}


\ccode{PACS Nos.: 95.35.+d, 95.36.+x, 95.75.-z, 98.35.-a, 98.62.-g, 98.80.-k}

\section{Introduction to Cosmological Structure Formation Simulations}

In the standard cosmological framework, the two most dominant components in the universe are dark energy (DE) and cold dark matter (CDM). The nature of dark energy is unknown and in the standard model it is assumed that it is a vacuum contribution in the form of a cosmological constant (generally denoted by $\Lambda$) in Einstein's field equation of general relativity.\cite{2001LRR.....4....1C, 2008PhLB..667..212P} Also the nature of the CDM is not understood and it could consist of weakly interacting massive particles (WIMPs) that only interact via the weak nuclear force and gravity. A possible WIMP candidate is the lightest supersymmetric particle (LSP) that is predicted in supersymmetric models of particle physics.\cite{1996PhR...267..195J, 2005PhR...405..279B, 2008PhLB..667..212P} Cold means that the thermal velocities of the CDM particles are small which leads to the formation of tiny structures in the universe.\cite{2004MNRAS.353L..23G, 2006PhRvL..97c1301P} The baryonic matter, of which all the stars and planets are made of, contributes only ca. 17\% of the total matter content.\cite{2009ApJS..180..330K} This standard cosmological framework is also known as the concordance or $\Lambda$CDM model. 

The parameters of the $\Lambda$CDM model are known to a good precision.\cite{2007ApJS..170..377S, 2009ApJS..180..330K} This leads to well defined initial conditions for cosmological structure formation simulations, which are completely determined by these parameters. The initial conditions are set up at early cosmological times when the density perturbations, which probably originate from quantum mechanical fluctuations in the inflationary epoch,\cite{2008PhLB..667..212P} are still in the linear regime. 

In most cosmological structure formation simulations, the baryonic matter is replaced by CDM due to computational reasons. The baryonic physics is complex and difficult to model in a self consistent way, which makes more realistic simulations very expensive. But the replacement is physically well motivated since for the large scale clustering properties of the CDM halos the baryonic contribution can be neglected. The baryons are only important at the centers of CDM halos: baryons can dissipate energy through radiative processes, hence they concentrate at the centers of the potential wells formed by the CDM halos where they can lead to a significant alteration of the CDM distribution. 

The non-linear evolution of the CDM is modelled as a collisionless fluid that interacts via gravity in an expanding universe. Collisionless means that the CDM can be described by a continuous distribution function that obeys the collisionless Boltzmann equation. The collisionless Boltzmann equation is a consequence of mass conservation under a Hamiltonian flow. The gravitational potential is then calculated via a Monte Carlo sampling of the continuous CDM distribution function by $N$ particles/bodies (sampling points). Hence this technique is called the $N$-body method.\cite{2008gady.book.....B} The sampling of the continuous distribution function introduces artificial singularities that are cut-off by softening the singular point particle potential at a softening scale $\epsilon$.\cite{2001MNRAS.324..273D}

In the last few years, an enormous effort in our collaboration went into understanding the structure of CDM halos of the size of our own galaxy, the Milky Way. In several large simulation projects, we put all the effort in resolving one such object within a larger cosmological volume, with a $\Lambda$CDM cosmology, with as much particles as computationally possible. This is achieved via a multi-mass nested refinement technique.\cite{1993ApJ...412..455K,2001ApJS..137....1B} The first simulation in this series was the Via Lactea I simulation,\cite{2007ApJ...657..262D, 2007ApJ...667..859D, 2007ApJ...669..676S, 2007ApJ...671.1135K, 2008ApJ...679.1260M, 2008ApJ...680L..25D, 2008AN....329.1004G, 2009ApJ...702..890G} which was followed by Via Lactea II.\cite{2008Natur.454..735D, 2009MNRAS.394..641Z, 2008ApJ...686..262K, 2008ApJ...689L..41M, 2009PhRvD..79l3517K, 2009PhRvD..80c5023B, 2009arXiv0905.4744M, 2009arXiv0906.1822K, 2009Sci...325..970K} Finally, the GHALO project was the first simulation to resolve a Milky Way sized CDM halo with more than one billion particles.\cite{2009MNRAS.398L..21S} The Aquarius project is a similar simulation project that was done by a different group.\cite{2008Natur.456...73S, 2008MNRAS.391.1685S, 2009MNRAS.395..797V, 2008arXiv0810.1522N} Aquarius has an equivalent resolution to GHALO. In this brief review, we will mainly focus on results from our projects.

\section{Numerical Aspects}

\subsection{Initial conditions}

The initial conditions of Via Lactea I \& II and GHALO are based upon the Wilkinson Microwave Anisotropy Probe (WMAP) three-year data\cite{2007ApJS..170..377S} and the three simulations have very similar cosmological parameters. In general, the initial conditions for cosmological structure formation simulations consist of a cubic volume of side length $L_\mathrm{Box}$ (see Table \ref{tab:summary}) with periodic boundary conditions. 

Creating initial conditions with billions of particles is a real challenge. In the case of GHALO, the GRAFIC1 and GRAFIC2 packages,\cite{2001ApJS..137....1B} which are often used by the community to produce initial conditions,\footnote{The original version of GRAFIC2 contained a bug that lead to incorrect initial conditions. Via Lactea II and GHALO used the corrected version of the code and are not affected by this. For more details consult Ref. \refcite{2009MNRAS.398L..21S}.} had to be parallelized.\cite{2009MNRAS.398L..21S} The initial conditions are then set up at high enough redshifts $z_\mathrm{IC}$ (see Table \ref{tab:summary}) when the density perturbations are still in the linear regime. The softening lengths $\epsilon$, where the singular point particle potential is cut-off, is also given in Table \ref{tab:summary}.

\subsection{Time evolution}

The consequent time evolution is done by numerically solving the coupled collisionless Boltzmann and Poisson equations. In the case of Via Lactea I \& II and GHALO this was done with the code PKDGRAV.\cite{2001PhDT........21S, 2008JPhCS.125a2008K} The latest version of PKDGRAV is a tree code that uses a fast multipole method (FMM) in Cartesian coordinates in order to calculate the gravity.\cite{2000ApJ...536L..39D, 2002JCoPh.179...27D} PKDGRAV uses a 5th-order reduced expansion in potential for faster and more accurate force calculation in parallel, and a multipole based Ewald summation technique for periodic boundary conditions. \citep{2001PhDT........21S} Much better parallel computing efficiency was achieved by using Single Instruction Multiple Data (SIMD) vector processing.

For the time integration, PKDGRAV uses a kick-drift-kick leapfrog integration scheme which is a second order, explicit, time-reversible and symplectic integrator when used with constant time-steps.\cite{2004shd..book.....L} The individual time-steps of the particles are chosen according to their local dynamical time which leads to a more accurate orbit integration in the centers of CDM halos and an overall more optimal time-step distribution which as a consequence requires less force evaluations.\cite{2007MNRAS.376..273Z}

\begin{table*}[t]
	\centering
	\tbl{Summary table for Via Lactea I \& II and GHALO. In the first block, basic halo properties at redshift $z=0$ are presented. The definitions of $r_\mathrm{200b}$, $M_\mathrm{200b}$, $r_{V_\mathrm{max}}$, $V_\mathrm{max}$ and $c_\mathrm{V}$ are given in the main text. The second block summarises some simulation properties: $m_\mathrm{p}$ is the particle mass in the high resolution region, $N_\mathrm{200b}$ is the number of particles within $r_\mathrm{200b}$, $z_\mathrm{IC}$ is the redshift of the initial conditions, $\epsilon$ is the softening length and $L_\mathrm{Box}$ is the size of the cubic cosmological box with periodic boundary conditions.}{
	\begin{tabular}{llccc}
		\toprule
		Simulation & & Via Lactea I & Via Lactea II & GHALO \\
		\colrule
		$r_\mathrm{200b}$      & [kpc]                  & 389 & 402 & 349 \\
		$M_\mathrm{200b}$      & [$\mathrm{M}_\odot$]   & $1.77 \times 10^{12}$ & $1.92 \times 10^{12}$ & $1.27 \times 10^{12}$ \\
		$r_{V_\mathrm{max}}$   & [kpc]                  & 69.0 & 59.8 & 49.5 \\
		$V_\mathrm{max}$       & [km s$^{-1}$]          & 181  & 201  & 153  \\
		$c_\mathrm{V}$         & [1]                    & $2.59 \times 10^{3}$ & $4.25 \times 10^{3}$ & $3.52 \times 10^{3}$ \\
		\colrule
		$m_\mathrm{p}$    & [$\mathrm{M}_\odot$] & $2.09 \times 10^{4}$ & $4.10 \times 10^{3}$ & $9.94 \times 10^{2}$ \\
		$N_\mathrm{200b}$ & [1]                  & $8.47 \times 10^{7}$ & $4.68 \times 10^{8}$ & $1.27 \times 10^{9}$ \\
		$z_\mathrm{IC}$   & [1]                  & 48.4 & 104 & 57.6\\
		$\epsilon$        & [pc]                 & 90 & 40 & 61\\
		$L_\mathrm{Box}$  & [Mpc]                & 90 & 40 & 40\\
		\botrule
	\end{tabular}
	\label{tab:summary}}
\end{table*}

\section{CDM Halos}\label{sec:general}

\subsection{General description}

In a $\Lambda$CDM cosmology, the main host halos contain a wealth of subhalos - smaller bound structures similar to their host halos. These subhalos contain again subsubhalos.\cite{2008Natur.454..735D, 2009MNRAS.398L..21S, 2008MNRAS.391.1685S} This is like a fractal pattern where a similar structure repeats on different scales which is expected to hold down to the cut-off scale of CDM. We denote the different levels of subhalos as sub$^n$halos, where $n$ can specify a particular level. 

\begin{figure}[t]
	\centering
	\includegraphics[width=\textwidth]{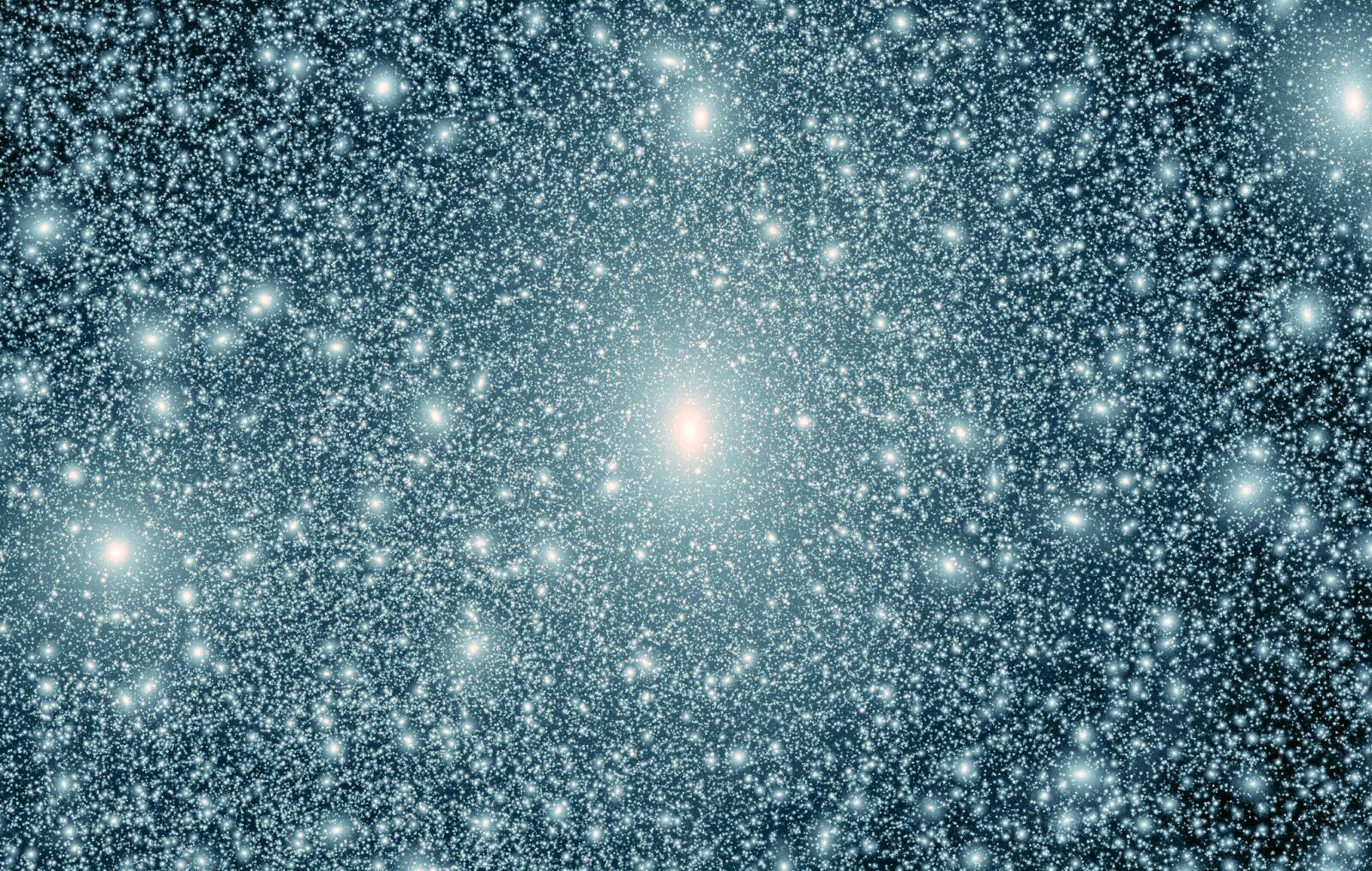}
	\caption{Dark matter mass density map of GHALO. The height of the picture corresponds to 400 kpc. The main halo, which is resolved with more than 1 billion particles, and the over $10^5$ numerically resolved subhalos it contains are visible as bright peaks.}
	\label{fig:ghalo}
\end{figure}

In Fig. \ref{fig:ghalo} we show the dark matter mass density map of GHALO. The main halo and the over $10^5$ numerically resolved subhalos it contains are visible as bright peaks. While smaller halos are falling towards a larger host halo, they get tidally stripped and loose mass. This lost mass is then detectable in phase space as coherent tidal streams that are streaming of the Lagrange points of the subhalo (see also Sec. \ref{sec:phasespace}).\cite{2008Natur.454..735D, 2009MNRAS.394..641Z} This tidal mass loss is not complete though and the inner, very compact cores of these infalling halos survive as sub$^n$halos within the host halo.

\subsection{Basic properties}

Defining the size of a CDM halo is a non-trivial task. The traditional method was based on the simple picture of the spherical collapse model.\cite{1972ApJ...176....1G} This model predicts, that after virialisation the virial radius $r_\mathrm{vir}$ encloses a region that has an overdensity of exactly $18\pi^2 (\approx 178)$ times the critical density $\rho_\mathrm{crit} = \frac{3 H^2}{8 \pi G}$ \footnote{$H$ is the Hubble parameter and $G$ is Newton's gravity constant.} in an Einstein-de Sitter cosmology. For the $\Lambda$CDM model the overdensity is approximately 100 with respect to $\rho_\mathrm{crit}$.\cite{1996MNRAS.282..263E, 1998ApJ...495...80B} But CDM halos in structure formation simulations collapse in a different way and the region that is dynamically influenced by the halo is much larger than the traditional virial radius.\cite{2006ApJ...645.1001P, 2007ApJ...667..859D, 2008ApJ...680L..25D} 

In this brief review we use the radius $r_{\mathrm{200b}}$ to characterise the size of CDM halos. The region within $r_{\mathrm{200b}}$ has an overdensity of 200 times the background, mean matter density $\rho_\mathrm{M}$. The associated enclosed mass is denoted by $M_{\mathrm{200b}} \equiv M(r_{\mathrm{200b}})$ where $M(r)$ indicates the enclosed mass. 

Due to the shape of the mass density profiles of CDM halos (see also Sec. \ref{sec:profile}), they have a well defined peak in the circular velocity function
\begin{equation}\label{eq:vcirc}
V_\mathrm{circ}(r) \equiv \sqrt{\frac{G M(r)}{r}}~.
\end{equation}
The peak value $V_\mathrm{max}$ of the circular velocity function $V_\mathrm{circ}(r)$ serves as proxy for mass and the radius $r_{V_\mathrm{max}}$, where the peak velocity $V_\mathrm{max}$ is reached, gives an inherent size scale of the CDM halo. Describing CDM halo properties by these two intrinsic scales avoids the trouble of defining an outer boundary of halos which is of especial advantage when characterising sub$^n$halos. The main properties of Via Lactea I \& II and GHALO are summarised in Table \ref{tab:summary}.

\section{Mass Distribution}

\subsection{Mass density profile}\label{sec:profile}

The simplest way of characterising CDM halos is by measuring their spherically averaged mass density profile. Over a large range of CDM halo masses, their mass density profiles can be well fit by a generalised NFW\cite{1996ApJ...462..563N} (GNFW) profile given by
\begin{equation}
\rho_\mathrm{GNFW}(r) \equiv \frac{\rho_s}{(r/r_s)^{\gamma} (1+ r/r_s)^{3-\gamma}}~.
\end{equation}
This form of the profile is a special case of the $\alpha\beta\gamma$ profile with $\alpha=1$ and $\beta=3$.\cite{1990ApJ...356..359H,1996MNRAS.278..488Z} In the outer part, the GNFW profile is $\propto r^{-3}$ whereas in the inner part $\propto r^{-\gamma}$, $r_s$ is the scale radius and $\rho_s$ the scale density. 

\begin{figure}[t]
	\centering
	\includegraphics[width=0.495\textwidth]{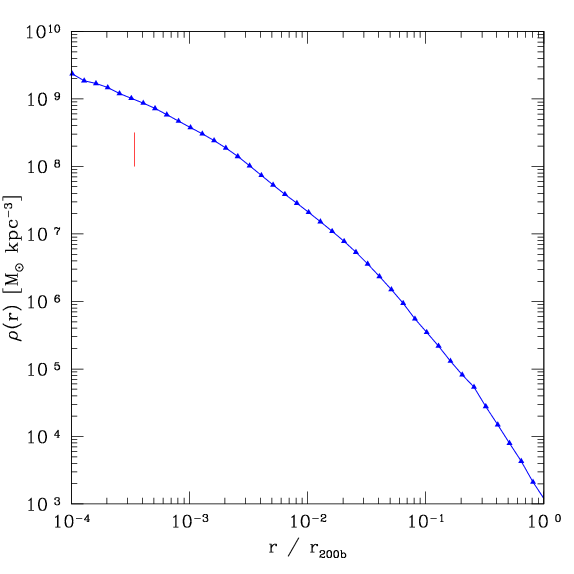}
	\includegraphics[width=0.495\textwidth]{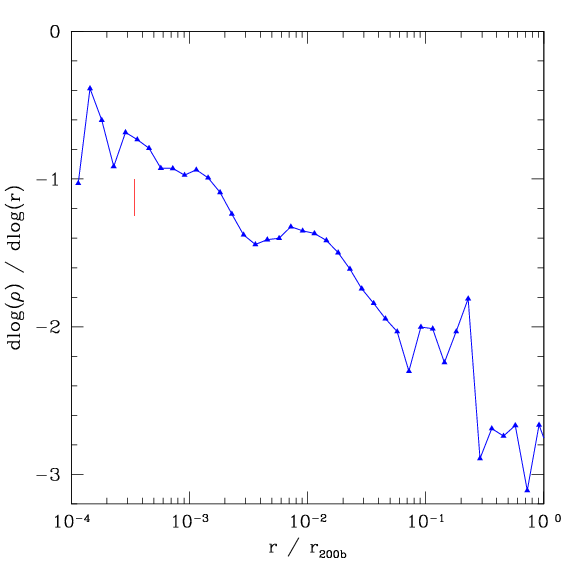}
	\caption{Mass density profile (left) and slope (right) for GHALO as a function of radius. For GHALO $r_\mathrm{200b}= 349~\mathrm{kpc}$. The vertical line marks the convergence radius of 120 pc.}
	\label{fig:rhoslope}
\end{figure}

The inner slope $\gamma$ was proposed to be $\gamma=1$, the NFW case, over a range of radii from $0.01~r_{\mathrm{200b}}$ to $r_{\mathrm{200b}}$.\cite{1996ApJ...462..563N} But it was soon realised, that CDM halos have some significant halo-to-halo scatter and most halos are significantly denser in their inner parts and have a steeper slope with $\gamma > 1$.\cite{1998ApJ...499L...5M, 1998MNRAS.300..146G, 2001ApJ...557..533F, 2003ApJ...588..674F, 2004ApJ...606..625F, 2005MNRAS.364..665D, 2008Natur.454..735D, 2009MNRAS.398L..21S} This is still true for our highest resolution case, GHALO, which shows a steep slope of -1.4 down to approximately $2\times10^{-3}~r_\mathrm{200b}$ (see Fig. \ref{fig:rhoslope}). 

Below this scale the mass density profile becomes significantly flatter due to the higher resolution and a newly proposed functional form for the mass density profile, the SM profile, 
\begin{equation}\label{eq:smprofile}
\rho_\mathrm{SM}(r) = \rho_0 e^{-\lambda[\ln(1+r/R_\lambda)]^{2}}
\end{equation}
fits better.\cite{2009MNRAS.398L..21S} The SM profile approaches the central maximum density $\rho_0$ as $r \rightarrow 0$. We also note that the derivative $\mathrm{d} \ln \rho/\mathrm{d} \ln(1 + r/R_\lambda)$ versus $\ln(1 + r/R_\lambda)$ has a slope of exactly -2$\lambda$. In the innermost region that is still resolved by GHALO, the logarithmic slope is -0.8 (see Fig. \ref{fig:rhoslope}).\cite{2009MNRAS.398L..21S} Such a shallow inner logarithmic slope on similar scales is also found in the Aquarius simulation.\cite{2008arXiv0810.1522N}

Locally, when measured in small volumes, the mass density can vary by orders of magnitudes in the outskirts of CDM halos. This is to some degree due to the prolate shape of the halos (see Sec. \ref{sec:shape}) but also because of the presence of both subhaloes and underdense regions or holes in the matter distribution.\cite{2009MNRAS.394..641Z} In the inner region, the dark matter distribution is very smooth and the contribution of subhalos is small.\cite{2009MNRAS.394..641Z, 2009MNRAS.395..797V}

Unfortunately, subhalos are still not resolved with enough particles in order to give a definitive answer about the shape of their mass density profile. But results from Via Lactea II show that they are also showing a steep inner cusp so that they would be similar to their host halos.\cite{2008Natur.454..735D}

\subsection{Concentration}

An intrinsic and general measure for the concentration of halos is given by the enclosed density within $r_{V_\mathrm{max}}$ in units of the critical density,\cite{2002ApJ...572...34A, 2007ApJ...667..859D}
\begin{equation}
c_\mathrm{V} \equiv \frac{\bar{\rho}(r_{V_\mathrm{max}})}{\rho_{\mathrm{crit},0}} = \frac{2}{H_0^2} \left(\frac{V_\mathrm{max}}{r_{V_\mathrm{max}}}\right)^{2}~.
\end{equation}
This definition has the advantage that it is also well defined for isolated halos as well as sub$^n$halos. The traditional virial concentration
\begin{equation}
c_\mathrm{vir} \equiv r_\mathrm{vir} / r_s
\end{equation}
depends on the detailed definition of the virial radius, the profile fitting procedure and has the drawback that it artificially increases due to the growth of $r_\mathrm{vir}$ with time.\cite{2007ApJ...667..859D} The usage of the inherent concentration $c_\mathrm{V}$ circumvents these complications and offers a robust concentration measure of halos as long as the peak of the circular velocity curve is resolved. The concentrations $c_\mathrm{V}$ for the host halos of Via Lactea I \& II and GHALO are also given in Table \ref{tab:summary}.

For isolated CDM halos of a given mass, objects that form earlier have a higher concentration. But there is a significant scatter in this relation.\cite{2001MNRAS.321..559B} Since in a $\Lambda$CDM cosmology smaller halos form in average earlier than larger halos, they have a higher concentration.\cite{2001MNRAS.321..559B, 2008MNRAS.391.1940M} Also, isolated halos in dense environments are more concentrated and exhibit a larger scatter than halos in a low density environment.\cite{2001MNRAS.321..559B}

Generally, subhalos have higher concentrations than isolated halos.\cite{2001MNRAS.321..559B} Subhalos that are located closer to the host halo center have higher concentrations than subhalos in the outskirts of the halo.\cite{2007ApJ...667..859D, 2008Natur.454..735D} This is due to the tidal forces that remove material of the infalling subhalos from the outer, low density region. Subhalos closer to the host halo center loose more mass due to the stronger tidal forces and more pericenter passages which leads to an increased concentration.\cite{2001ApJ...563....9M, 2007ApJ...667..859D}

\subsection{Shape}\label{sec:shape}

The mass distribution of isolated CDM halos has the shape of triaxial ellipsoids in general, though the shape is close to prolate since the lengths of the intermediate axis $b$ and the minor axis $c$ are similar.\cite{2006MNRAS.367.1781A, 2007MNRAS.376..215B, 2007ApJ...671.1135K, 2009MNRAS.398L..21S} Less massive halos are in general rounder, i.e. the ratio $c/a$, where $a$ is the major axis length, is larger for less massive halos.\cite{2006MNRAS.367.1781A, 2007MNRAS.376..215B} 

Overall, GHALO has a prolate shape as well with $b/a \approx c/a \approx 0.5$. The outer parts of GHALO, and CDM halos in general, are rounder and the axis ratios become larger with distance. At the halo centers the shape diverges fast and becomes more spherical due to numerical reasons.\cite{2009MNRAS.398L..21S}

Subhalos have a triaxial shape in general as well but are over all more spherical than the host halos. Like isolated halos, subhalos tend to be less spherical in their central regions. Also, subhalos are tendentially slightly more spherical closer to the host halo center.\cite{2007ApJ...671.1135K}

\section{Subhalos}

\subsection{Spatial distribution and orientation of subhalos}

When halos fall into a host halo and become subhalos, their mass is removed from outside inwards by tidal interactions. This process leads to an antibiased spatial distribution of subhalos with respect to the smooth background mass and subhalos have a more extended spatial distribution. The tidal disruption process is not able  to completely disrupt the subhalos though and the dense cores survive. For example in Via Lactea I, 97\% of all subhalos at redshift $z=1$ survive until the present time and the total mass in resolved subhalos is 5 \% at redshift $z=0$.\cite{2007ApJ...667..859D, 2008ApJ...679.1260M} The volume occupied by subhalos increases with distance from the center of the host halo.\cite{2009MNRAS.394..641Z}

When selecting subhalos by their current mass, they have a number density distribution at redshift $z=0$ of
\begin{equation}
n_\mathrm{sub,M}(r) \propto \rho(r) ~ r
\end{equation}
where $\rho$ is the smooth DM density.\cite{2004MNRAS.352..535D, 2007ApJ...667..859D, 2008MNRAS.391.1685S} The subhalo number density is independent of the selection mass threshold and the subhalo mass, i.e. subhalos in different mass bins have the same spatial distribution when properly normalised.\cite{2004MNRAS.352..535D, 2008MNRAS.391.1685S, 2009ApJ...692..931L} $V_\mathrm{max}$ is less affected by tidal mass removal and therefore $V_\mathrm{max}$ selected subhalos follow closer the smooth DM distribution at redshift $z=0$ with
\begin{equation}
n_\mathrm{sub,V_{max}}(r) \propto \rho(r) ~ M(r)
\end{equation}
where $M(r)$ denotes the enclosed mass.\cite{2007ApJ...667..859D, 2008Natur.454..735D, 2008MNRAS.391.1685S} 

During their orbit within the host halos, the major axis of subhalos tends to be aligned with the direction toward the host halo center. This is effect is most likely due to tidal interactions as well in the sense that tides stretch the subhalo shape along the direction towards the center. For subhalos further out (i.e. distances larger than 3 $r_\mathrm{200b}$), this alignment is not observed any more.\cite{2007ApJ...671.1135K}

\subsection{Sub$^n$halo abundance}

The cumulative subhalo abundance within $r_\mathrm{200b}$ when selected by velocity (velocity function) at redshift $z=0$ can be described by
\begin{equation}
N(>V_\mathrm{max}) \propto \left(\frac{V_\mathrm{max}}{V_\mathrm{max,host}}\right)^{-\alpha_V}
\end{equation}
with $\alpha_V = 3$ and where $V_\mathrm{max,host}$ is the peak velocity of the host halo.\cite{1998MNRAS.300..146G, 2007ApJ...667..859D, 2008Natur.454..735D, 2008MNRAS.391.1685S} For Via Lactea II the proportionality factor is 0.036 and $V_\mathrm{max,host} = 201~\mathrm{km}~\mathrm{}^{-1}$ (see also Table \ref{tab:summary}).\cite{2008Natur.454..735D} The velocity function is independent of the selection radius, i.e. in the case of Via Lactea II the velocity functions of subhalos selected within 100 kpc or 50 kpc have the same shape as the velocity function for subhalos selected within $r_\mathrm{200b}$, only the normalisation changes.\cite{2008Natur.454..735D} The mean sub$^2$halo abundance is consistent with a scaled down version of the main halo within the current resolution of simulations.\cite{2008Natur.454..735D}

When selecting by mass, the cumulative subhalo abundance (mass function) at redshift $z=0$ is well described by 
\begin{equation}
N(>M) \propto \left(\frac{M}{M_\mathrm{200b,host}}\right)^{-\alpha_M}
\end{equation}
with $\alpha_M = 0.9 \ldots 1.0$.\cite{1998MNRAS.300..146G, 1998ApJ...499L...5M, 2007ApJ...657..262D, 2007ApJ...667..859D, 2008MNRAS.391.1685S}

\section{Velocity Space Structure}

\subsection{Anisotropy parameter}

A parameter to describe the velocity space structure of CDM halos is the anisotropy parameter
\begin{equation}
\beta \equiv 1 - \frac{1}{2}\frac{\sigma_\mathrm{tan}^2}{\sigma_\mathrm{rad}^2}
\end{equation}
where $\sigma_\mathrm{tan}$ is the tangential velocity dispersion and $\sigma_\mathrm{rad}$ the radial velocity dispersion. A value of $\beta \approx 0$ means isotropic, $\beta > 0$ respectively $\beta < 0$ denotes a radial respectively tangential anisotropy. For relaxed CDM halos $\beta(r)$, measured in spherical shells, changes from nearly isotropic in the center to slightly radial in the outskirts of the halo.\cite{1996MNRAS.281..716C, 2000ApJ...539..561C, 2001ApJ...557..533F, 2004MNRAS.351..237R, 2004MNRAS.352..535D, 2006NewA...11..333H, 2009MNRAS.394..641Z}

When $\beta$ is measured in small, local volumes, then the local value can greatly deviate from the spherically averaged value. In the inner halo, regions along the long shape axis particles are preferentially on radial orbits whereas along the short shape axis particles are on tangential orbits. This effect disappears at larger radii.\cite{2009MNRAS.394..641Z}

\subsection{Velocity dispersion ellipsoid}

As a consequence of the tensor virial theorem, the global velocity dispersion ellipsoid is aligned with the shape ellipsoid.\cite{2008gady.book.....B} This is also true for small local volumes in the inner parts of halos, where the local velocity ellipsoid is aligned with the global shape as well. In the outskirts this local alignment is not observed any more.\cite{2009MNRAS.394..641Z}

The shape of the global velocity dispersion ellipsoid (and as a consequence the potential) is significantly rounder than the mass distribution.\cite{2007ApJ...671.1135K} When measured again in small local volumes, the shape depends on the location within the halo.\cite{2009MNRAS.394..641Z}

\subsection{Velocity distribution function}

\begin{figure}[t]
	\centering
	\includegraphics[width=0.495\textwidth]{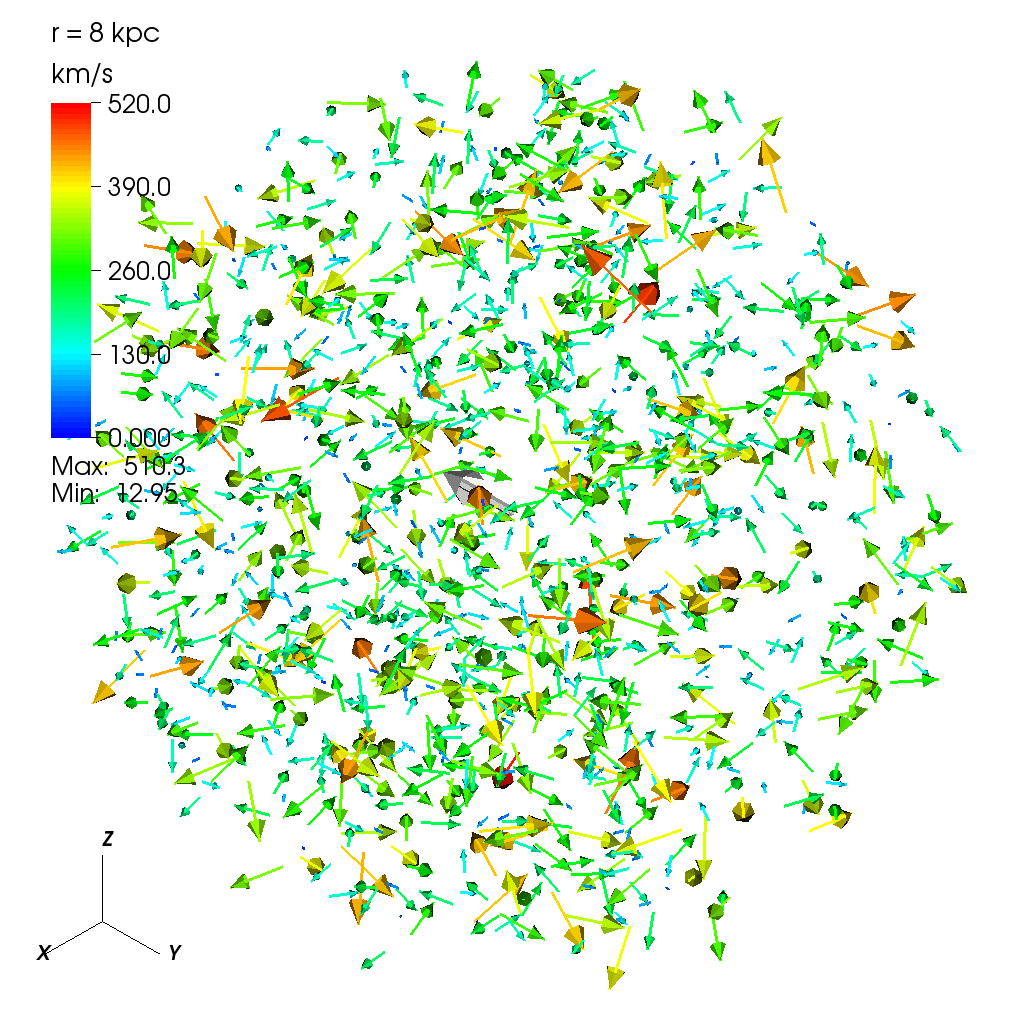}
	\includegraphics[width=0.495\textwidth]{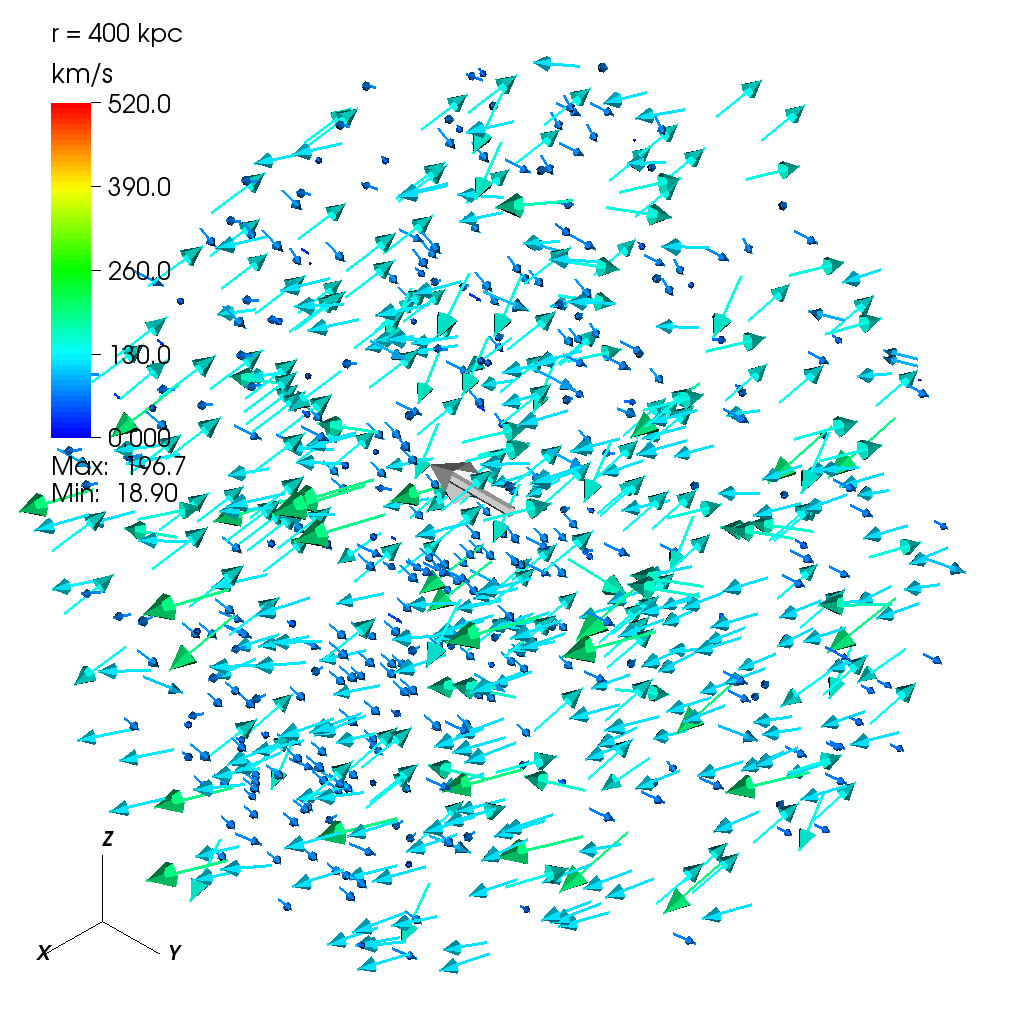}\\
	\includegraphics[width=0.495\textwidth]{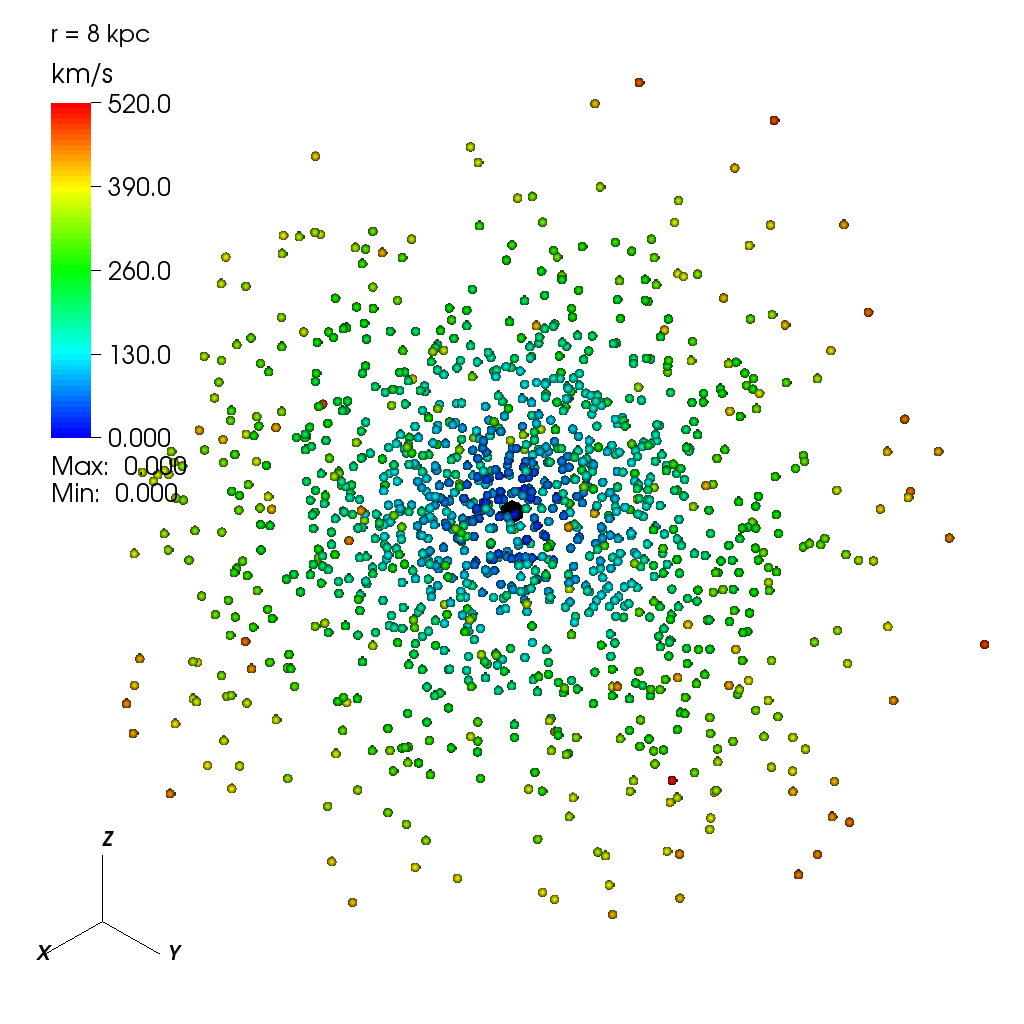}
	\includegraphics[width=0.495\textwidth]{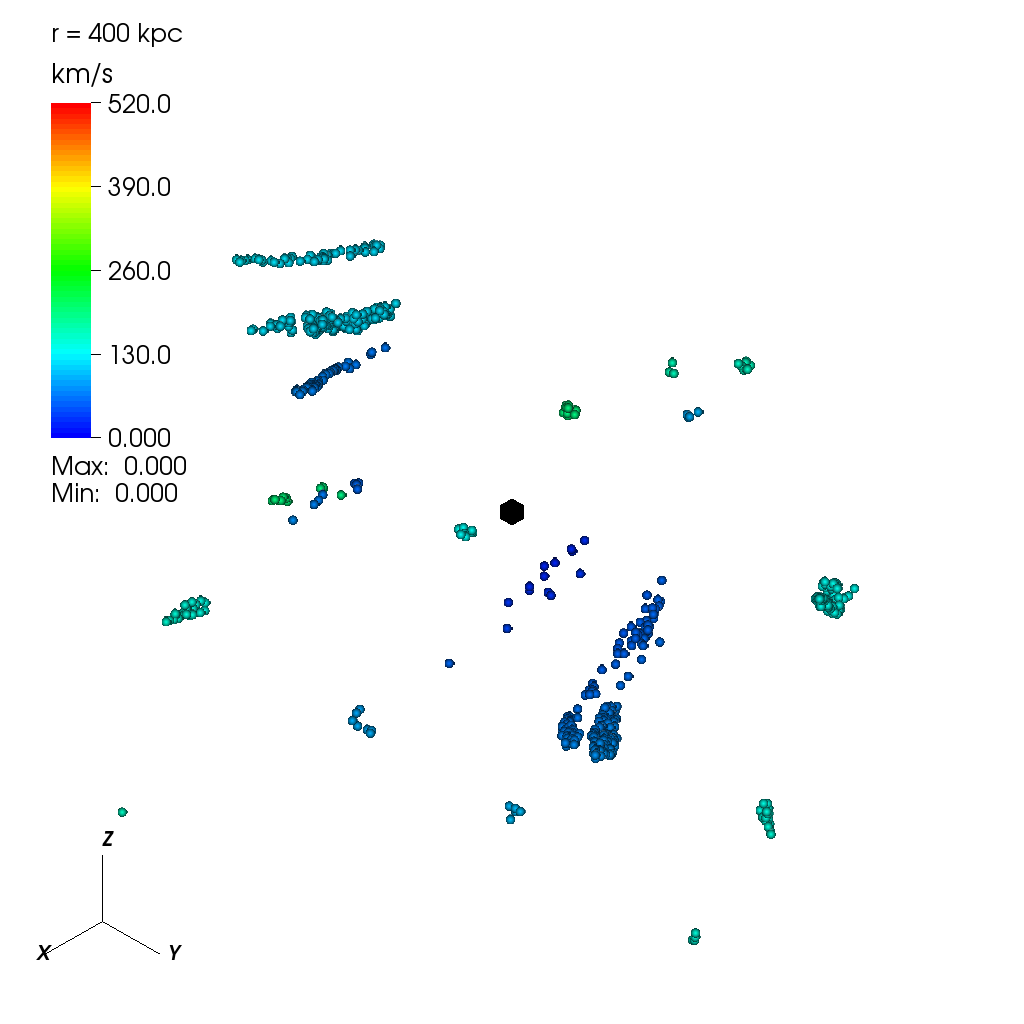}
	\caption{Velocity structure of Via Lactea II. Positions (top row) of particles within small volumes at 8 kpc (left) and 400 kpc (right) from the Via Lactea II center as well as the corresponding location of the particles in the local velocity space (bottom row). The colour in both cases is given by the magnitude of the velocity vector.}
	\label{fig:localvelocity}
\end{figure}

From theoretical models that describe CDM halos with smooth phase space distribution functions, it is expected that the local velocity distribution has the shape of a normal distribution around regions where the mass density profile has a slope of -2, i.e $\rho \propto r^{-2}$. In the inner regions, where the mass profile is shallower, the velocity distribution is more peaked and in the outer region, where the mass profile is steeper, the velocity distribution is broader.\cite{1990ApJ...356..359H, 2004ApJ...601...37K, 2008MNRAS.388..815W} 

But a smooth velocity distribution is only a good description of the velocity space of CDM halos in the central region. In the outer parts, the local matter stems from several tidal streams of subhalos that lost material while passing. This is best illustrated in Fig. \ref{fig:localvelocity}. This figure shows the positions and velocity vectors (top row) for every particle within local volumes centered at 8 kpc (left) and 400 kpc (right) from the center of Via Lactea II. We also plot the location of the particles in the local velocity space (bottom row). The colour (and length in position space) encodes the magnitude of the velocity vector. The big white arrow in the center of the position space plots points towards the galactic center whereas the black cube in the center of the local velocity space plots marks the origin. For the velocity space plots, we split the velocity vector field in a radial- ($x$-axis), $\varphi$- ($y$-axis) and a $\vartheta$-component ($z$-axis).

These small volumes have been selected to be free of subhalos. Therefore, one would not expect to see a clumpy structure in the velocity space of these volumes. At 8 kpc no velocity space structure is apparent. The orientation of the velocity vectors appears random and the actual velocity distribution can be well fit by a multivariate normal distribution. But at 400 kpc, there is evidence for locally coherent motion, visible as groups of vectors with the same colour pointing in the same direction. This is even clearer in the velocity space plot: there are only a bit more than a dozen clumps in velocity space and no smooth component at all. Obviously a smooth multivariate normal distribution would not provide a good fit. This is evidence that the outer part of the halo is built up from a collection of large scale streams from the tidal disruption of infalling subhaloes. But so too is the smooth component in center, in that it also likely consists of an overlap of many, many streams.\citep{2003MNRAS.339..834H, 2009arXiv0906.4341V} The central limit theorem then guarantees that the resulting distribution in the center closely resembles something like a multivariate normal distribution\citep{2003MNRAS.339..834H} or more general distributions.\citep{2006JCAP...01..014H, 2009MNRAS.395..797V}

\section{Phase Space Structure}\label{sec:phasespace}

\begin{figure}[t]
	\centering
	\includegraphics[width=0.495\textwidth]{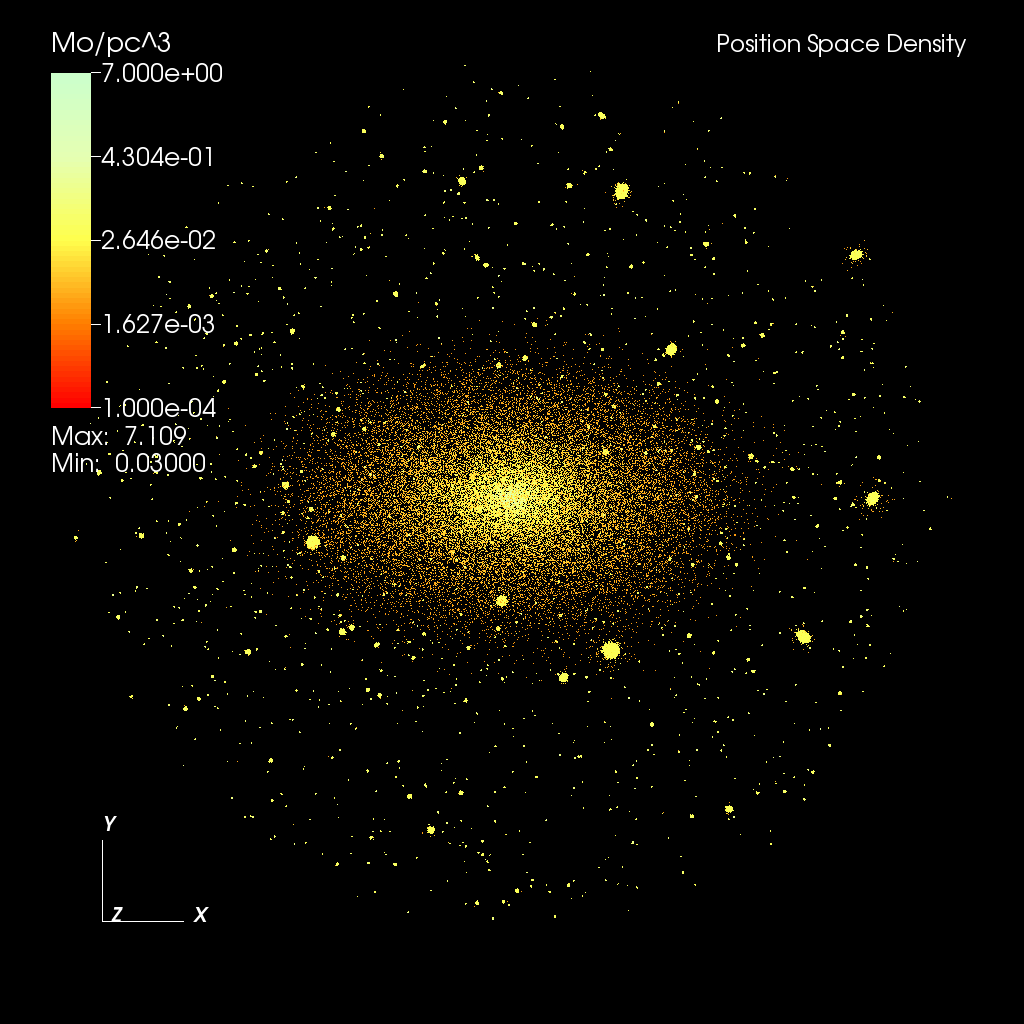}
	\includegraphics[width=0.495\textwidth]{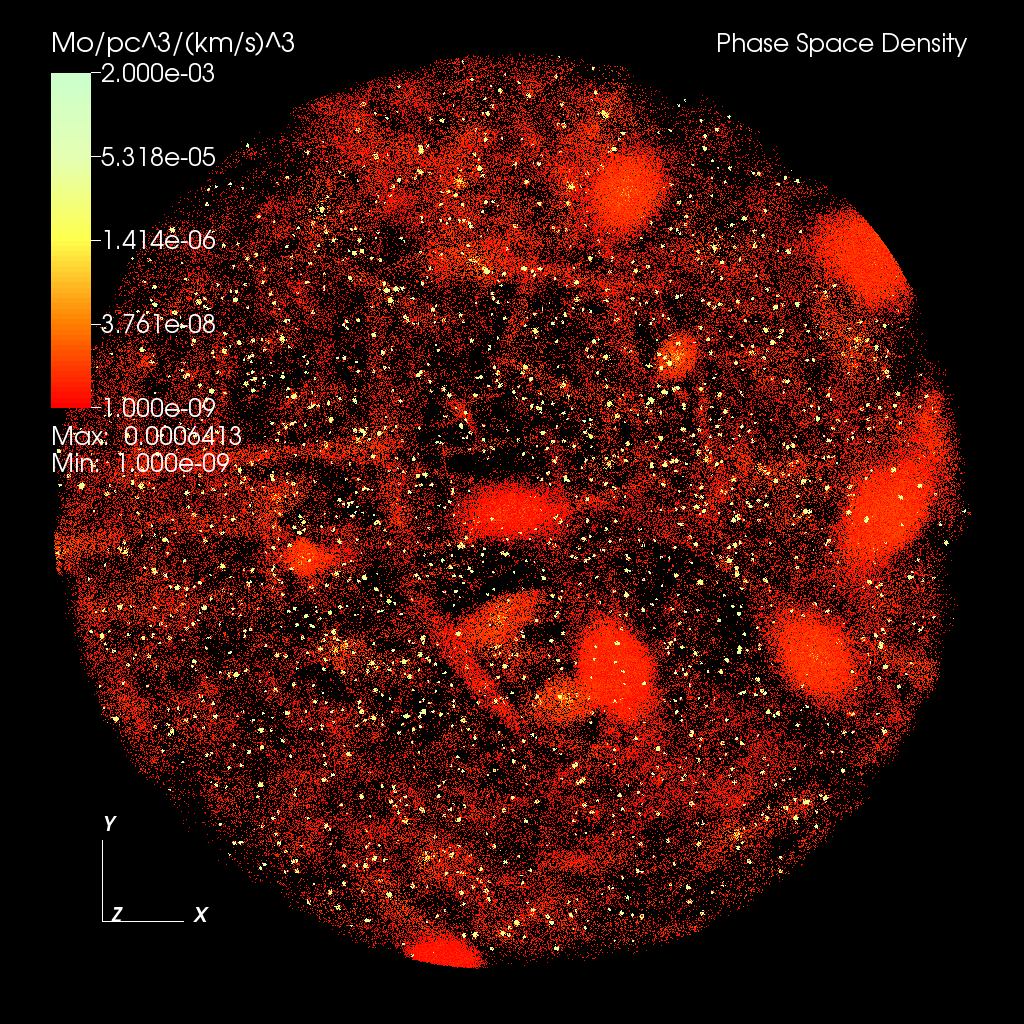}
	\caption{Central sphere of radius 50 kpc where we calculated the position space density (left) and the true phase space density (right) with EnBiD. Approximately 2000 peaks from the subhaloes within that radius are visible in both pictures but the contrast of subhaloes is higher in the phase space density picture. In the phase space picture are also large scale dark matter streams visible which were formed by tidal mass loss of infalling subhaloes. These streams are not visible in the traditional position space density picture.}
	\label{fig:denpsd}
\end{figure}

In Fig. \ref{fig:denpsd} we show the central region of Via Lactea II: the sphere has a radius of 50 kpc. We plot the position space density (left) and the true six dimensional phase space density (right) calculated with EnBiD\footnote{\texttt{http://sourceforge.net/projects/enbid/}}.\citep{2006MNRAS.373.1293S} We only show the top five to six orders of magnitude in position and phase space density, in order not to overload the two pictures. Many cold streams are clearly visible in the central region. Although these cold streams only contribute a few percent to the local mass density, their velocity dispersion is just a few km s$^{-1}$, resulting in a very high phase space density for these particles. These streams are not visible in traditional density picture (see Fig. \ref{fig:ghalo}) and can only be revealed by visualising the true phase space density. Via Lactea II seems to be the first structure formation simulation with sufficient resolution to reveal these streams in phase space, since previous lower resolution runs did not show such phase space features. Also subhaloes are better visible due to their higher contrast in phase space density. Approximately 2000 peaks are seen in the central 50 kpc sphere phase space density image. The high contrast makes the phase space density the ideal method of identifying subhaloes.

An often used method of describing the structure of phase space is by the pseudo phase space density $Q \equiv \rho/\sigma^3$ where $\rho$ is the matter density and $\sigma$ is the velocity dispersion. When $Q$ is measured in spherical shells, then it follows a power law
\begin{equation}
Q(r) \propto r^{-\alpha_Q}
\end{equation}
with $\alpha_Q = 1.8 \ldots 1.95$.\cite{2001ApJ...563..483T, 2004MNRAS.351..237R, 2004MNRAS.352.1109A, 2005MNRAS.363.1057D, 2008MNRAS.386.2022A, 2009MNRAS.398L..21S}

\begin{figure}[t]
	\centering
	\includegraphics[width=0.5\textwidth]{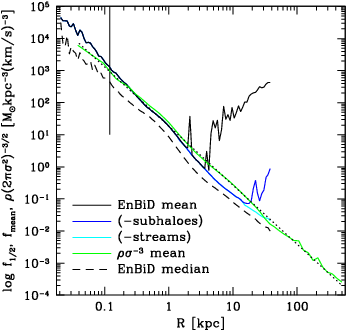}
	\caption{Phase space density profile for GHALO. The $Q(r)/(2\pi)^{3/2}$ profile follows a power law with slope -1.84 (dotted line). The mean of the true phase space density calculated with EnBiD is dominated by the contribution from subhalos beyond 1.8 kpc. The approximate power law relation can be recovered by removing high phase space density contributions from subhalos and streams. The true phase space density relation is steeper and has a slope of ca. -2. The median of the true phase space density is not affected by the contribution from substructure. The resolution limit of GHALO is 120 pc and is marked with a vertical line. This figure was adopted from Ref. \protect \refcite{2009MNRAS.398L..21S}.}
	\label{fig:psd}
\end{figure}

In Fig. \ref{fig:psd} we plot the quantity $Q(r)/(2\pi)^{3/2}$,\footnote{The additional factor $(2\pi)^{3/2}$ is necessary in order to agree with the expectation value from a Maxwell-Boltzmann distribution.} which follows a nice power law with a slope of $\alpha_Q = 1.84$ for GHALO.\cite{2009MNRAS.398L..21S} It is interesting to compare this to the spherically averaged true phase space density again calculated with EnBiD. When taking the mean true phase space density in spherical shells, we see that subhalos start to dominate the profile from ca. 1.8 kpc on where the innermost subhalo in GHALO is located. In the inner part, the mean EnBiD value roughly shows a power law behaviour but with a steeper slope of -2. Subhalos have in general a higher central phase space density than the host halos.\cite{2004MNRAS.353...15A, 2009MNRAS.395.1225V} By cutting out the contribution of particles with $f > 100 ~\mathrm{M}_\odot ~\mathrm{kpc}^{-3} ~\mathrm{km}^{-3} ~\mathrm{s}^{3}$ we can remove the contribution from subhalos. When removing also streams by neglecting particles with $f > 0.4~\mathrm{M}_\odot ~\mathrm{kpc}^{-3} ~\mathrm{km}^{-3} ~\mathrm{s}^{3}$ the approximate power law behaviour can be extended out to ca. 40 kpc. Also, the median of the true phase space density is not affected by the contribution form substructure (see Fig. \ref{fig:psd}).

The power law behaviour of the pseudo phase space density $Q$ is still enigmatic and likely reflects the distribution of entropy $K \equiv Q^{-2/3} = \sigma^2/\rho^{2/3}$, which dark matter acquires as it is accreted.\cite{2009MNRAS.395.1225V} Therefore, one should use $Q$ with caution and give preference to the true phase space density when studying the phase space structure.

\section{Conclusions and Final Remarks}

In this brief review we discuss recent results from the high resolution structure formation simulations Via Lactea I \& II and GHALO which study the formation of Milky Way sized objects. The resulting CDM halos show a wealth of structure: several levels of sub$^n$halos and streams in phase space in regions that where smooth in previous simulations. We present quantitative descriptions of the mass and subhalo distribution as well as a detailed insight into the velocity and phase space structure.

Unfortunately, the content of this brief review is limited and we had to skip results on implications for indirect and direct dark matter detection experiments. Please consult Refs. \refcite{2007ApJ...657..262D,2008ApJ...686..262K,2009PhRvD..79l3517K,2009PhRvD..80c5023B,2009arXiv0906.1822K,2009Sci...325..970K} for more results in that area. For a complementary review on the structure and evolution of CDM halos consult Ref. \refcite{2009arXiv0906.4340D}. 

Images, movies and data from Via Lactea I \& II and GHALO are available at \texttt{http://www.ucolick.org/$\sim$diemand/vl/} and \texttt{http://www.itp.uzh.ch/ghalo/}.

\section*{Acknowledgements}

It is a pleasure to thank my collaborators J{\"u}rg Diemand, Michael Kuhlen, Piero Madau, Ben Moore, Doug Potter, Joachim Stadel and Larry Widrow for useful discussions and input. M.Z. is supported by NSF grant AST-0708087.

\bibliography{RDB_ISO}

\end{document}